\documentclass[twocolumn,showpacs,preprintnumbers,amsmath,amssymb]{revtex4}
\usepackage{graphicx}
\usepackage{dcolumn}
\usepackage{bm}

\begin{document}

\title{Oxygen Vacancy Induced Ferromagnetism in V$_2$O$_{5-x}$}

\author{Zhi Ren Xiao}
\author{Guang Yu Guo}\email{Electronic address: gyguo@phys.ntu.edu.tw}
\affiliation{Department of Physics and Center for Theoretical Sciences, National Taiwan University, Taipei 106, Taiwan}
\author{Po Han Lee}
\affiliation{The Affiliated Senior High School of National Taiwan Normal University, Taipei 106, Taiwan}
\author{Hua Shu Hsu}
\author{Jung Chun Andrew Huang}
\affiliation{Department of Physics, National Cheng Kung University, Tainan 701, Taiwan}

\date{\today}

\begin{abstract}
{\it Ab initio} calculations within density functional theory with
generalized gradient approximation have been performed to study the effects
of oxygen vacancies on
the electronic structure and magnetism in undoped V$_2$O$_{5-x}$ ($0 < x < 0.5$).
It is found that the introduction of oxygen vacancies would induce ferromagnetism
in V$_2$O$_{5-x}$ with the magnetization being proportional to the O
vacancy concentration $x$. The calculated electronic structure
reveals that the valence electrons released by the introduction of oxygen
vacancies would occupy mainly the neighboring V $d_{xy}$-dominant band which then becomes
spin-polarized due to intra-atomic exchange interaction, thereby giving rise to the
half-metallic ferromagnetism.
\end{abstract}

\maketitle


Ferromagnetism in diluted magnetic semiconductors (DMSs) has recently received a lot
of attention for fundamental research and possible applications in
spintronics\cite{ohn98,pri98,das01}. Although a number of transition-metal
(TM) oxide based DMSs, such as Co:TiO2 and Co:ZnO, have been
reported,\cite{gri05,ued04} the origin of ferromagnetism in many cases
is still under much debate.\cite{wen04,kun04,ued01}
The controversy of the magnetic origin lies in the structure of
DMSs in the first place, i.e., whether it is due to the intrinsic contribution
from the DMS phase or from some secondary phases.\cite{gri05}  It is noticed
that even if the formation of the secondary phases could
be ruled out, the valence effect on ferromagnetism, the role of the microscopic complex structure
formed by oxygen, dopant, and host TM elements need to be clarified.
Oxygen vacancies may also play an important role for the ferromagnetism in
oxide based DMSs because oxygen vacancies not only modulate the valence
of neighboring TM elements but also cause a change
of the band structure of host oxides; both factors in turn can make significant contribution
to the ferromagnetism.\cite{coe05} Several conflicting experimental results of oxide based DMSs
could be partly related to the difference in their oxygen vacancies because of distinct
sample growth conditions.\cite{gri05,coe05}.
Therefore, understanding of the role of oxygen vacancies in TM oxides is essential to clarify
the origin of the ferromagnetism in the oxide based DMSs.

Vanadium oxides consist of compounds with single [V$_2$O$_5$ (V$^{+5}$),
V$_2$O$_4$ (V$^{+4}$), V$_2$O$_3$ (V$^{+3}$) and VO (V$^{+2}$)]
and mixed valences which form a fascinating class of multifunctional materials.
Vanadium oxides have been widely used in catalytic, optical, electrical and
electrochemical devices.\cite{suc05,cha05,raj05}
V$_2$O$_5$ is a diamagnetic insulator at room temperature
with an orthorhombic layered structure. It undergoes a phase transformation to
a metallic phase at about 530 K.\cite{yoo05,ram97} By comparison, V$_2$O$_4$
is a diamagnetic semiconductor with a tetragonal rutile structure below a critical
temperature of $\sim$340 K and a metal with a monoclinic structure above.\cite{eye02}.
While the physical properties for various vanadium oxides have been extensively
studied,\cite{kel04,bao97,de04} the role of oxygen vacancies on the structural,
electronic and magnetic properties of V$_2$O$_5$ remains yet to be explored.
We have therefore performed {\it ab initio} theoretical studies of the electronic
structure and magnetism in V$_2$O$_{5-x}$. Our calculations reveal that the presence
of oxygen vacancies introduces electrons onto the neighboring V $d_{xy}$ dominant
conduction band which is then spin splitted due to
intraatomic exchange interaction, leading to the formation of the magnetic order.
Our finding calls for critical re-examinations of the
role of the intrinsic O vacancies in the formation of magnetism in TM
based DMSs, and also suggests a new
direction for search for ferromagnetism in semiconductors.

Our {\it ab initio} calculations are based on density functional theory with the
generalized gradient approximation to the electron exchange-correlation
interaction.\cite{per96}. Two approaches are used to model the oxygen vacancies
in V$_2$O$_{5-x}$. One is the so-called virtual crystal approximation (VCA).
In V$_2$O$_{5-x}$, $x$ O atoms are removed, and this is equivalent to increase
the number of the conduction electrons by 2$x$. In the VCA, this is simulated 
by artificially increasing the electron and atomic number of V from 23.0 to
$23.0 + x$, i.e., replacing the V atoms by the virtual atoms with
an atomic number of $23.0 + x$.
We have performed self-consistent band structure calculations for V$_2$O$_{5-x}$ with
several $x$ values in both nonmagnetic (NM) and ferromagnetic (FM) states.
We use the highly accurate full-potential linearized augmented plane wave (FLAPW) method,
as implemented in WIEN2k package.\cite{bla01} The experimental lattice constants used here are $a$ = 11.512 \AA,
$b$ = 3.564 \AA, and $c$ = 4.368 \AA.~\cite{enh86} The cutoff angular momentum for the spherical harmonics
expansion is 10 and as many as $\sim$120 augmented plane waves per atom are included.

\begin{figure}[t]
\begin{center}
\end{center}
\caption{(Color) Isosurface of the spin density
of V$_2$O$_{5-x}$ ($x = 0.125$) in the 1$\times$2$\times$2 supercell.
The numbers are the numberings of the V atoms in the unit cell
that are refered to in the text and Table I.}
\label{f1}
\end{figure}

The other approach is the supercell (SC) method which is much more CPU time-consuming especially
for small $x$ values. Nonetheless, the SC method is more realistic.
In particular, we can study local structural, electronic
and magnetic properties in the vicinity of the O vacancies. We have carried the SC calculations
for the $x = 0.25$ and $0.125$ cases using the $1\times1\times2$ and $1\times2\times2$ supercells,
respectively. Pure V$_2$O$_5$ has a layered structure with two formula
units (f.u.) per unit cell.\cite{enh86} Each layer contains edge- and corner-shared VO$_5$ pyramids with
three types of O atoms, namely, apical oxygen (O1), chain oxygen (O2) and bridge oxygen (O3),
as illustrated in Fig. 1. In order to investigate which kind of the O vacancies
is energetically favorable, we perform {\it ab initio} structural optimizations using
the $1\times 1\times 2$ supercell with one O1 or O2 or O3 vacancy, respectively.
We use the computationally more efficient projector augmented wave (PAW) method,\cite{blo94}
as implemented in VASP package.\cite{kre96}
A large cutoff energy of 400 eV is used, and the equilibrium atomic positions are obtained
when the forces acting on all the atoms were less than 0.02 eV/\AA.
Our $1\times1\times2$ SC calculations show that the total energy
per vacancy for the O1 vacancy is 1.41 and 2.15 eV lower than that of the O2 and O3 vacancies,
respectively, suggesting that apical oxygen (O1) is the preferred O vacancy site.
Thus, we perform the structural optimization for the $1\times2\times2$ SC with only
the O1 vacancy. Our $1\times1\times2$ and $1\times2\times2$ SC calculations show
that the O1 vacancy in the FM state has a lower total energy than in the NM state,
respsectively, by 78 and 86 meV per vacancy, showing that the
ferromagnetism is stable in V$_2$O$_{5-x}$. The stable FM state
is also found in the VCA calculations for all the $x$ values considered.
We also perform the self-consistent spin polarized
calculations for the antiferromagnetic (AF) configuration with the initial magnetic moment on
the V atom next to the vacancy being antiparallel to the other V atoms. However,
the self-consistent calculations always converge to the FM state, indicating that
the AF state is unstable.

Final self-consistent band structures for the $1\times2\times2$ supercell
with the theoretically determined atomic
structure were calculated using the FLAPW method.
The calculated V local magnetic moments for $x = 0.125$ are listed in
Table I. The total magnetic moment is 2.0 $\mu_B$ per vacancy, and is the same as
that obtained from the VCA calculations.
Majority of the magnetization is located on the V atoms ($d_{xy}$-orbital) next to
the O vacancy (V7,V3,V9 in Fig.1) and also
in the surrounding interstitial region, as can also be seen from
Fig. 1 which shows the spin density distribution.
The induced magnetization on all the O atoms are negative (i.e., antiparallel to that
on the V atoms) and the sum of the O atomic moments is -0.360 $\mu_B$ per supercell. 
However, the local magnetic moments on $most$ O atoms are negligible ($<0.01 \mu_B$), 
though six O atoms have a magnetic moment of $0.01\sim 0.05 \mu_B$.

\begin{figure}[t]
\begin{center}
\includegraphics[width=8cm]{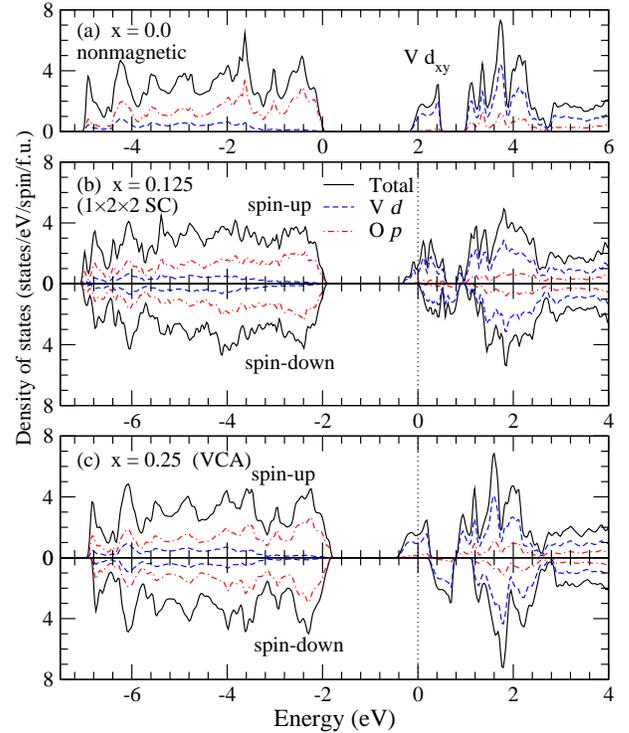}
\end{center}
\caption{Total and site-decomposed density of states for
V$_2$O$_{5-x}$ with (a) $x=0.0$, (b) $x=0.125$ (SC calculation) and
(c) $x=0.25$ (VCA calculation).}
\label{f2}
\end{figure}

Fig. 2(a) shows that pure V$_2$O$_5$ is a semiconductor with a band gap of $\sim$1.8 eV.
While the valence band is clearly O $p$ dominant, the lowest conduction band is of mainly
V $d_{xy}$ character. Interestingly, Fig. 2(b)
shows that when one O vacancy is introduced, the lowest conduction band is strongly spin-polarized,
and the spin up one is partially occupied whilst the spin down one is still empty, resulting in
a FM state with a spin moment of 2 $\mu_B$ per vacancy.
Clearly,
when one O atom is removed, the two electrons released are added to the narrow V $d_{xy}$ dominant
conduction band. However, because of the rather strong V intra-atomic exchange
interaction ($I_{ex}$)
of $\sim$0.354 eV,\cite{jan77} this partially occupied V $d$ band becomes magnetically
unstable and spin split, giving
rise to a ferromagnetic order, as the spin-polarized band structure for $x = 0.125$
and $0.25$ in Fig. 2(b-c) shows.
To corroborate this Stoner mechanism for ferromagnetism, we also calculate
total and site-decomposed density of states in the nonmagnetic V$_2$O$_{5-x}$ ($x = 0.125$) in the $1\times2\times2$ supercell. Table I shows that V7 next to the
O vacancy (see Fig. 1) has such a large local density of states at the Fermi level
($N(E_F)$) that the Stoner criterion ($I_{ex}N(E_F) > 1$) for the magnetization
formation is satisfied.
Fig. 3 shows the calculated magnetic moment per V atom as function of O vacancy
concentration ($x$). Remarkably, the calculated magnetic moment per V atom are proportional to $x$.
This is due to the fact that the additional 2$x$ electrons released by removing $x$
O vacancies go exclusively to occupy the lowest V $d$-conduction band of the
majority spin, as mentioned above.

\begin{table}[t]
\caption{Local magnetic moments ($m_s$) ($\mu_B$/atom) of V atoms
in the ferromagnetic V$_2$O$_{5-x}$ ($x$ = 0.125) from
the $1\times 2\times 2$ supercell
calculation. The total magnetic moment is 2.0 $\mu_B$
per O vacancy. The locations of various V atoms are indicated in Fig. 1.
Int. denotes the magnetization in the interstitial region.
Also listed are the V-site decomposed
density of states at the Fermi level $N(E_F)$ (states/eV/atom)
in the nonmagnetic V$_2$O$_{5-x}$.}
\label{t1}
\begin{center}
\begin{tabular}{@{\hspace{\tabcolsep}\extracolsep{\fill}}cccccc} \hline
 V             &  $m_s$ & $N(E_F)$&    V          &  $m_s$ & $N(E_F)$\\ \hline
 1 ($\times$2) &  0.049  & 0.43   & 8             & 0.073  & 1.29  \\
 2 ($\times$2) &  0.046  & 0.72   & 9             & 0.267  & 1.36  \\
 3             &  0.278  & 3.47   & 10            & 0.118  & 0.57 \\
 4             &  0.075  & 1.15   & 11 ($\times$2)& 0.039  & 0.37 \\
 5             &  0.177  & 0.69   & 12 ($\times$2)& 0.030  & 0.49 \\
 6             &  0.087  & 0.46   & int.       &    0.595  & 1.99\\
 7             &  0.360  & 4.35   &            &           &  \\ \hline
\end{tabular}
\end{center}
\medskip
\end{table}

\begin{figure}[t]
\begin{center}
\includegraphics[width=8cm]{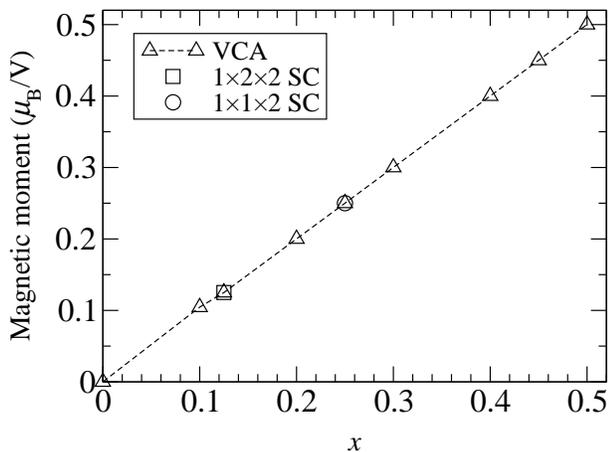}
\end{center}
\caption{Magnetic moment per V atom as a function
of O vacancy concentration $x$ from VCA and SC calculations.}
\label{f3}
\end{figure}


Summarizing, our {\it ab initio} calculations show that V$_2$O$_{5-x}$,
with $0.0 < x < 0.5$ is ferromagnetic. The two valence electrons released by the introduction of each
oxygen vacancy would occupy mostly the $d_{xy}$ orbital of the V atoms next to the vacancy, and become
fully spin-polarized due to intra-atomic exchange interaction, giving rise to the ferromagnetic
order with a magnetic moment of 2$\mu_B$ per vacancy. We note that recent experiments indicate
possible room-temperature ferromagnetism in oxygen-deficient V$_2$O$_{5-x}$~\cite{hua06}.
We also note that the recent {\it ab initio}
calculations indicate the unexpected ferromagnetism observed in HfO$_2$ being also driven
by intrinsic O vacancies.\cite{das05}
Nevertheless, the microscopic mechanisms in the present and previous cases would be fundamentally
different since in Ref. \cite{das05}, the magnetic moment is centered at the O vacancy.
Clearly, the present work calls for critical re-examinations of the magnetism in TM oxide
based DMSs especially the role of the intrinsic O vacancies. The present work
also suggests a new direction for search for ferromagnetism in semiconductors.

The authors gratefully acknowledge financial supports from National Science Council
of Taiwan. They also thank National Center for High-performance
Computing of Taiwan for providing CPU time.\\


\begin{thebibliography}{9}

\bibitem{ohn98} H. Ohno: Science {\bf 281} (1998) 951.

\bibitem{pri98} G. A. Prinz: Science {\bf 282} (1998) 1660.

\bibitem{das01} S. Das Sarma: Am. Sci. {\bf 89} (2001) 516.

\bibitem{gri05} K. A. Griffin, A. B. Pakhomov, C. M. Wang, S. M. Heald, and K. M. Krishman:
 Phys. Rev. Lett. {\bf 94} (2005) 157204.

\bibitem{ued04} K. Ueda, H. Tabata, and T. Kawai: Phys. Rev. Lett. {\bf 93} (2004) 177206.

\bibitem{wen04} H. Weng, X. Yang, J. Dong, H. Mizuseki, M. Kawasaki, and Y. Kawazoe:
 Phys. Rev. B {\bf 69} (2004) 125219.

\bibitem{kun04} D. C. Kundayiya, S. B. Ogale, S. E. Lofland, S. Dhar, C. J. Metting, S. R. Shinde,
Z. Ma, B. Varughese, K. V. Ramanujachary, L. Salamanca-Riba and T. Venkatesan:
Nature Mater. {\bf 3} (2004) 709.

\bibitem{ued01} K. Ueda, H. Tabata, and T. Kawai: Appl. Phys. Lett. {\bf 79} (2001) 988.

\bibitem{coe05} J. M. D. Coey, M. Venkatesan and C. B. Fitzgerald: Nature Mater. {\bf 4} (2005) 173.

\bibitem{suc05} Y. Suchorski, L. Rihko-Struckmann, F. Klose, Y. Ye, M. Alandjiyska, K. Sundmacher and
H. Weiss: Appl. Surf. Sci. {\bf 249} (2005) 231.

\bibitem{cha05} Y. J. Chang, C. H. Koo, J. S. Yang, Y. S. Kim, D. H. Kim, J. S. Lee, T. W. Noh,
H.-T. Kim, and B. G. Chae: Thin Solid Films {\bf 486} (2005) 46.

\bibitem{raj05} B. Rajesh, K. Ravindranathan Thampi, J.-M. Bonard, H. J. Mathieu, N. Xanthopoulos
and B. Viswanathan: J. Power Sources {\bf 141} (2005) 35.

\bibitem{yoo05} M. H. Yoon, and S. Im: Appl. Surf. Sci. {\bf 244} (2005) 444.

\bibitem{ram97} C. V. Ramana, O. M. Hussain, B. Srinivasulu Naidu, and P. J. Reddy:
Thin solid Films {\bf 305} (1997) 219.

\bibitem{eye02} V. Eyert: Ann. Phys. (Leipzig) {\bf 11} (2002) 650.

\bibitem{kel04} G. Keller, K. Held, V. Eyert, D. Vollhardt, and V. I. Anisimov:
 Phys Rev. B {\bf 70} (2004) 205116.

\bibitem{bao97} W. Bao, C. Broholm, G. Aeppli, P. Dai, J. M. Honig and P. Metcalf:
Phys. Rev. Lett. {\bf 78} (1997) 507.

\bibitem{de04} L. A. L. de Almeida, G. S. Deep, A. M. N. Lima, I. A. Khrebtov,
V. G. Malyarov, and H. Neff: Appl. Phys. Lett. {\bf 85} (2004) 3605.

\bibitem{per96} J. P. Perdew, K. Burke, and M. Ernzerhof: Phys. Rev. Lett. {\bf 70} (1996) 045422.

\bibitem{bla01} P. Blaha, K. Schwarz, G. K. H. Madsen, D. Kvasnicka, and J. Luitz,
{\it WIEN2K, An Augmented Plane Wave + Local Orbitals
Program for Calculating Crystal Properties} (K. Schwarz, Techn. Univ. Wien, Austria, 2001).

\bibitem{enh86} R. Enjalbert and J. Galy: Acta Crystallogr. C {\bf 42} (1986) 1467.

\bibitem{blo94} P. E. Bl\"{o}chl: Phys. Rev. B {\bf 50} (1994) 17953 (1994);
G. Kresse and J. Joubert: Phys. Rev. B {\bf 59} (1999) 1758.

\bibitem{kre96} G. Kresse and J. Furthm\"{u}ller: Comput. Mat. Sci. {\bf 6} (1996) 15 (1996);
Phys. Rev. B {\bf 54} (1996) 11169.

\bibitem{jan77} J. F. Janak: Phys. Rev. B {\bf 16} (1977) 255.

\bibitem{hua06} J. A. C. Huang (2006, unpublished)

\bibitem{das05} C. Das Pemmaraju and S. Sanvito: Phys. Rev. Lett. {\bf 94} (2005) 217205.

\end{thebibliography}
\end{document}